\documentclass[a4paper,11pt]{article}
\pdfoutput=1 

\usepackage{jheppub} 

\usepackage[T1]{fontenc} 

\usepackage{graphicx,subfigure}
\usepackage{amsmath,amssymb,amsfonts}
\usepackage{bm}
\usepackage{color}
\usepackage{lineno}
\usepackage{comment}
\usepackage{hyperref}

\def \be {\begin{equation} }
\def \ee {\end{equation}}
\def \bea {\begin{eqnarray}}
\def \eea {\end{eqnarray}}
\def \bem {\begin{multline}}
\def \eem {\end{multline}}

\def \bes {\begin{subequations} }
\def \ees {\end{subequations}}

\def \pd {\partial}

\def \a {\alpha}
\def \b {\beta}

\def \e {\epsilon}

\def \s {\sigma}

\def \<{\langle}
\def \>{\rangle}
\def \+{\dagger}
\def \({\left(}
\def \){\right)}

\def \[{\left[}
\def \]{\right]}

\def \vx {\bm{x}}
\def \vs {\bm{s}}

\def \vj {\bm{j}}

\def \vB {\bm{B}}

\def \vv {\bm{v}}

\def \CC {{\cal C}}

\def \CT {{\cal T}}
\def \CL {{\cal L}}
\def \CO {{\cal O}}

\def \tr {\tilde{r}}

\def \no {\nonumber}

\def \vAfun {V_{A}(r)}
\def \vanom {v_{\rm anom}}
\def \etaA {\eta_{\rm anom}}

\def \rca {r^{0}_{c}}
\def \gxx {g_{xx}(r)}

\def \rH {r_{H}}

\def \anom {{\rm anom}}

\def\drag {{\rm drag}}

\graphicspath{{./plots/}}

\begin{document}

\vspace{5mm}
\preprint{MIT-CTP/4727}

%
%

\title{\bf 
The charmonium dissociation in \\ an ``anomalous wind"
}

\author[a,b]{Andrey V. Sadofyev}
\author[c]{Yi Yin}
\affiliation[a]{Center for Theoretical Physics, Massachusetts Institute of Technology, Cambridge, MA 02139, USA}
\affiliation[b]{ITEP, B. Cheremushkinskaya 25, Moscow, 117218, Russia}
\affiliation[c]{Physics Department,
Brookhaven National Laboratory, Upton, NY 11973, USA}

\emailAdd{sadofyev@mit.edu}
\emailAdd{phyyin@gmail.com}

\abstract{ 
We study the charmonium dissociation in a strongly coupled chiral plasma in the presence of magnetic field and axial charge imbalance.
This type of plasma carries ``anomalous flow" induced by the chiral anomaly and exhibits novel transport phenomena such as chiral magnetic effect.
We found that the ``anomalous flow" would modify the charmonium color screening length by using the gauge/gravity correspondence.  
We derive an analytical expression quantifying the ``anomalous flow" experienced by a charmonium for a large class of chiral plasma with a gravity dual.
We elaborate on the similarity and {\it qualitative} difference between anomalous effects on the charmonium color screening length which are {\it model-dependent} and those on the heavy quark drag force which are fixed by the second law of thermodynamics. 
We speculate on the possible charmonium dissociation induced by the chiral anomaly in heavy ion collisions.
%
}

\date{\today}
\maketitle

\section{Introduction and summary}

The work of Matsui and Satz \cite{Matsui:1986dk} introduced the idea of using a quarkonium to probe quark gluon plasma (QGP). 
In a deconfined QGP, 
quarkonium bound states such as charmonium $c\bar{c}$ will dissociate because of color screening and thereby exhibit a suppression relative to the confined phase. 
Due to its importance,
the problem of charmonium dissociation in QGP has been a focus of many recent studies 
(see Refs.~\cite{Rapp:2008tf,CasalderreySolana:2011us} for reviews and references).

In this paper, we consider the problem of charomium dissociation in a chiral (parity-violating) plasma with a finite chiral (axial) charge density\footnote{For a recent discussion of lattice QCD with chiral chemical potential see e.g. \cite{Braguta:2015zta}} and in the presence of an external magnetic field. 
This environment is pertinent to the QGP, which is approximately chiral, created in heavy-ion collisions.
First, 
a very strong magnetic field is generated from the incoming nuclei that are positively charged and move at nearly the speed of light. 
Such magnetic field has a magnitude of the order of $eB\sim m^2_{\pi}$ and its lifetime can be significant when medium's effect is taken into consideration~\cite{Tuchin:2013ie,Gursoy:2014aka}. 
Meanwhile, 
QCD as a non-Abelian gauge theory has topologically nontrivial gluonic configurations such as instantons and sphalerons. 
These configurations couple to quarks through the chiral anomaly and translate topological fluctuations into the chiral imbalance for quarks.

The focus of our study is on the effects of the chiral anomaly on the color screening length $l_{s}$, which is an important parameter quantifying charmonium dissociation. 
In heavy-ion collisions, 
the produced charmonium is moving relative to QGP and the relative velocity $v$
 (or rapidity $\eta=\tanh^{-1}(v)$) can be significant. 
We therefore also take dependence of $l_{s}$ on rapidity $\eta$ into consideration. 

Anomaly-induced effects in a chiral medium has attracted much interests recently 
(see Refs.~\cite{Kharzeev:2012ph,Zakharov:2012vv,Kharzeev:2013ffa,Liao:2014ava} for reviews).
One familiar example is the chiral magnetic effect (CME) \cite{Vilenkin:1980fu,alekseev1998universality, Kharzeev:2004ey,Kharzeev:2007tn,Kharzeev:2007jp,Fukushima:2008xe}, 
the generation of a vector current $\vj_{V}$ along an external magnetic field $\vB$.
In particular and closely related to the current work, 
those anomalous effects modify hydrodynamics of chiral fluids \cite{Son:2009tf} 
(see also Refs.~\cite{Sadofyev:2010is,Neiman:2010zi,Sadofyev:2010pr}
). 
For such fluid in the frame that energy density is at rest (i.e. in the Landau frame), the entropy density is not at rest and is moving opposite to the direction of chiral magnetic current $\vj_{\rm CME}=C_{A}\mu_{A}\vB$ :
\be
\label{va_def}
\vs_{\anom}= s \vanom \hat{B}\, , 
\qquad
\vanom \equiv -\frac{C_{A}\mu_{A}\mu}{\(\e + p\) }B\, , 
\qquad
C_{A}=
\frac{N_c }{2\pi^{2}}\, , 
\ee
where $\e,~s,~p,~\mu~(\mu_{A})$ denotes the energy density, entropy density, pressure and vector (axial) chemical potential respectively and the coefficient $C_{A}$ is fixed by the chiral anomaly\footnote{We set charge $e=1$ here but will recover $e$-dependence in \eqref{CA_QCD} .} 
In this work, we will use charmonium (or in general quarkonium) to probe such an anomalous chiral fluid 
and ask how its screening length $l_{s}(\eta;\vanom)$ would be influenced by the presence of the ``anomalous flow" $\vanom$.

To compute the rapidity-dependent color screening length $l_{s}(\eta;\vanom)$, we will use the gauge/gravity correspondence following the general formalism of Ref.~\cite{Liu:2006nn}. 
Previously, $l_{s}(\eta)$ and the dissociation of a moving charmonium has been studied in the framework of holographic correspondence from both top-down \cite{Liu:2006nn,Caceres:2006ta} and bottom-up \cite{Hohler:2013vca} approaches. 
Quarkonia dissociation in the presence of magnetic field has also been addressed previously (see for example Refs.~\cite{Marasinghe:2011bt,Alford:2013jva,Dudal:2014jfa,Cho:2014loa,Guo:2015nsa}). To the extent of our knowledge,
the effects of the chiral anomaly on the charmonium dissociation have not been reported in literature before.

The main finding of this paper is that the charmonium color screening length $l_{s}(\eta;\vanom)$ receives contributions from the chiral anomaly: 
a charmonium finds itself in a wind induced by anomalous flow \eqref{va_def}. 
Let us quantify the ``anomalous flow" felt by a charmonium introducing $\eta_{\anom}$ with following properties:
\begin{equation}
\label{eta_anom}
l_{s}(\eta;\vanom)
=l_{s}(\eta+\eta_{\anom};\vanom=0)\, .
\end{equation}
In other words,
the color screening length of a charmonium moving at rapidity $\eta$ in the \textit{presence} of the anomalous flow 
$\vanom$ equals to that of a charmonium moving at rapidity $\eta+\eta_{\anom}$ in the \textit{absence} of the anomalous flow. 
For small $\vanom \ll 1$,  
we obtain an analytical expression within the current holographic model at linear order in $\vanom$ for $\eta_{\anom}$.
We observe that the magnitude of $\eta_{\anom}$ is proportional to $\vanom$. 
However, its value is {\it model-dependent}.
Very recently, anomalous contributions to the heavy quark drag force were studied for a holographic chiral fluids~\cite{Rajagopal:2015roa}. 
Our study here provides further insights on the ``anomalous flow" felt by heavy probes of the chiral plasma.

This paper is organized as follows. In Sec.~\ref{sec:set_up}, we describe our holographic set-up.  
In Sec.~\ref{sec:ls},
we derive the analytic formula which determines the anomalous contribution to the color screening length $l_{s}(\eta;\etaA)$.
At this point, our results are valid for a large class of holographic chiral fluids.
In Sec.~\ref{sec:results},
we take ${\cal N}=4$ SYM theory as an example and present $\eta_{\anom}$ as well as anomalous contributions to $l_{s}(\eta;\vanom)$. 
We compare our results with anomalous contributions to the heavy quark drag force found in Ref.~\cite{Rajagopal:2015roa,Stephanov:2015roa} and speculate on phenomenological implications for heavy-ion collisions in Sec.~\ref{sec:summary}.

\section{The holographic setup}
\label{sec:set_up}
\begin{figure}
\centering
\includegraphics[width=0.6\textwidth]
{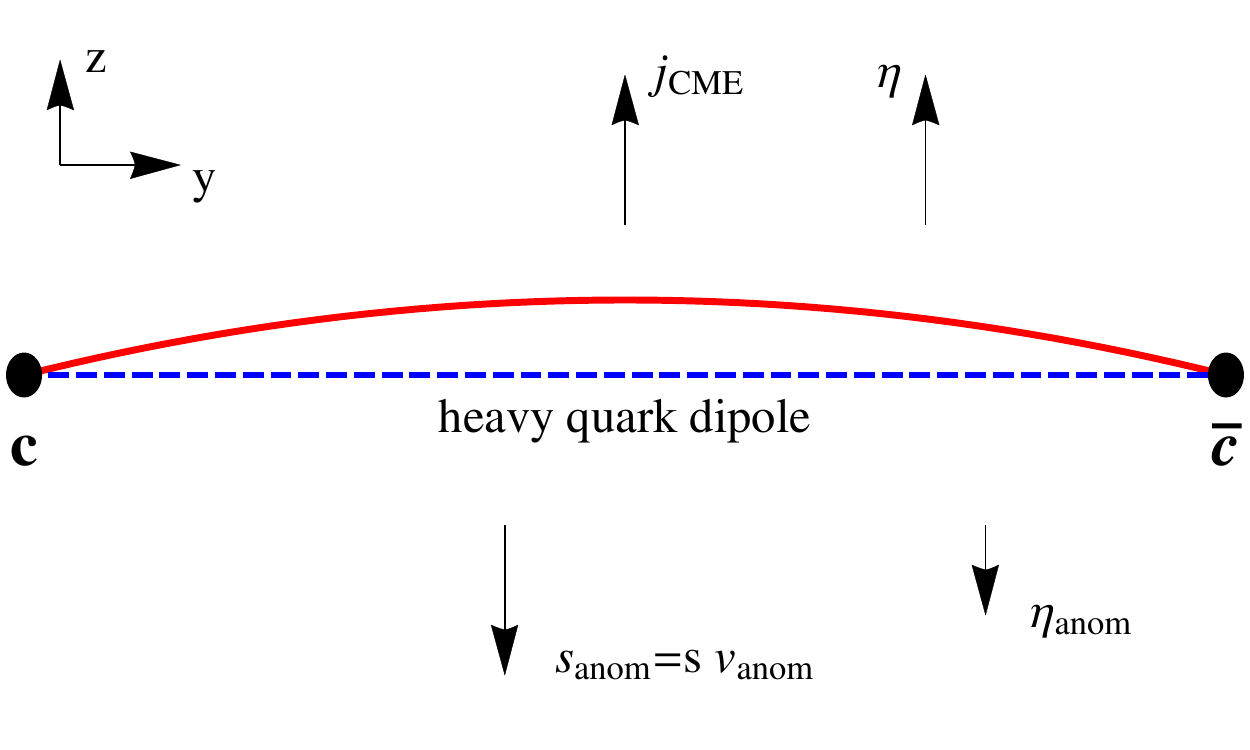}
\caption
{
\label{fig:profile}
(Color Online) 
A schematically view of the heavy-quark ``dipole" configuration and the corresponding string world-sheet (projected into $y-z$ plane) in the holographic set-up considered in this paper. 
The chiral magnetic current is along $z$-direction and the ``dipole" is placed along $y$-direction. 
The entropy flow $\vanom$ in \eqref{va_def} is opposite
to the chiral magnetic current.
We will show $\eta_{\anom}$ \eqref{eta_anom} is also along $\vanom$ direction. 
Here the ``dipole" is moving at rapidity $\eta$ along $\vanom$ direction.
Nevertheless, $\eta_{\anom}$ is invariant under $\eta \to -\eta$. 
Without the anomalous flow $\vanom$,
the string profile is constant along $z$-direction (dashed blue curve).
However, 
the anomalous flow will stretch the string profile along $z$-direction (red curve, see \eqref{sec:z1} for details). 
}
\end{figure}

We start with the $4+1$-dimensional asymptotic AdS bulk metric which is dual to the dense strongly coupled plasma of ${\cal N}=4$ SYM theory with a nonzero axial chemical potential (as a manifestation of the chiral imbalance) and with a homogeneous external magnetic field $\vB$.
Such metric can be determined by solving the bulk Einstein-Maxwell-Chern-Simons equations. 
 For analytical transparency, 
 we will consider the situation that $\vanom$ given by \eqref{va_def} is small. 
Consequently, the bulk metric, at leading order in $\vanom$, 
can be obtained by solving the linearized bulk equation of motion around AdS Reissner-Nordstrom (RN) black-brane solution, see Refs.~\cite{Erdmenger:2008rm,Banerjee:2008th,Son:2009tf} for details. 
The resulting metric,
in the Landau frame of the fluid, is of the form:
\be
\label{metric2}
d s^{2} =\gxx \{ - f(r) dt^2 + d\vx^2 -2
\vanom Q(r) dz dt\}   -2 \frac{\vanom Q(r)}{f(r)}dz dr
+\[\gxx f(r)\]^{-1}dr^2 \, .
\ee
Here $r$ denotes holographic coordinates with the boundary at $r\to \infty$ and the horizon at $r\to r_{H}$, where $f(\rH)=0$. 
Without losing generality, we take the chiral magnetic current to be along the $z$ direction (c.f.~ Fig.~\ref{fig:profile}). 

To keep our discussion as general as possible, we will defer writing down explicitly $r$-dependence of the metric \eqref{metric2} to Sec.~\ref{sec:results}.
Our discussion presented in this section and Sec.~\ref{sec:ls} can be applied to a large class of chiral fluids in the presence of the anomalous flow with holographic dual of the form \eqref{metric2}. 
 In \eqref{metric2},
 $g_{tz}(r)$ and $g_{zr}(r)$ which are proportional to $\vanom Q(r)$ are induced by the chiral anomaly.

We now consider a ``dipole" moving through a thermal plasma at the velocity $\vv$. 
To be specific, $\vv$ here denotes the velocity with respect to the Landau rest frame. 
To concentrate on the effects related to the chiral anomaly,  
we take the velocity $\vv$ to be along $z$ direction (see~ Fig.~\ref{fig:profile} for a schematic view). 
To proceed, it is convenient to consider the gravity background \eqref{metric2} in the frame that the ``dipole" is at rest while energy density of the plasma is moving. 
This metric can be determined by boosting  the metric \eqref{metric2} using:
\be
\label{boost}
dt= \cosh\eta \, dt'- \sinh\eta\, dz'\, , 
\qquad
dz=- \sinh\eta \, dt' +\cosh\eta\, dz'\, .
\ee 
Here we have also introduced the rapidity: 
\begin{equation}
v \equiv \tanh(\eta)\, .
\end{equation}
As a result, we have (dropping the prime to save notations):
\begin{eqnarray}
\label{metric2_v}
ds^2&=&
G_{tt}(r)dt^2 +G_{zz}(r)dz^2 +\gxx\(dx^2+dy^2\) -\(\frac{1}{\gxx f}\)dr^2 
\no \\
&+&2 \[ G_{tz}(r)dtdz -  \vanom \cosh\eta\, Q(r) f^{-1}dzdr+\vanom\sinh\eta\,  f^{-1}(r) Q(r) dt dr \], 
\end{eqnarray}
where:
\begin{eqnarray}
\label{G_metric}
G_{tt}(r) &=& \(\cosh\eta\)^2 \gxx \Big[\(- f(r)+\tanh^2\eta\)+2 \vanom\tanh\eta\,  Q(r) \Big] 
\, ,
\qquad \\
G_{zz}(r) &=& \(\cosh\eta\)^2\gxx\Big[\(1 - \tanh^2\eta f(r)\)+ 2 \vanom\tanh\eta\, Q(r)\Big]
\, , 
\no \\
G_{tz}(r)
&=& -\(\cosh\eta\)^2 \gxx \Big[ \tanh\eta\(1-f(r)\) + \vanom Q(r)\(1+\tanh^2\eta\)  \Big] 
\, .
\end{eqnarray}

To describe the interaction potential energy $E$ of a heavy quark ``dipole" with quarks separated by a distance $L$ and moving at rapidity $\eta$,
we consider the Wilson loop in the frame \eqref{metric2_v} whose contour $\CC$ is given by a rectangle with large extension $\CT$ in the $t$-direction and short sides of length $L$ along some spatial direction. 
 We take short sides of $\CC$ to lie in the $y$ direction that is transverse to both $\vv$ and ``anomalous flow" $\vv_{\anom}$. 
(c.f.~Fig.~\ref{fig:profile}). 
The generalization to the arbitrary angle between $L$ and $\vv$ is straightforward. 

Following holographic dictionary, 
the interaction potential energy $E$ of the ``dipole", measured by the thermal expectation value $\<W(\CC)\>$, is given by
\cite{Liu:2006he}:
\be
e^{-i E \CT}
= \<W(C)\> = e^{i S[\CC]}\, , 
\ee
where $S\[\CC\]$ is related to the corresponding Nambu-Goto action:
\be
\label{S_def}
S\[\CC\]
= \frac{1}{ 2 \pi \a}
\int d\s d\tau 
\CL 
= \frac{1}{ 2 \pi \a}
\int d\s d\tau 
\sqrt{-{\textrm det}\, h_{\a\b}}\, , 
\ee
with the induced metric given by:
\be
\label{h_def}
h_{\a\b}= G_{M N}\pd_{\a}X^{M}\pd_{\b}X^{N}\, .
\ee
Here, $M,N$ run over $t,\vx, r$ and $\a,\b$ run over $\tau, \s$ and $G_{MN}$ can be read from \eqref{metric2_v}. 

The action $S(\CC)$ is invariant under the choice of $\tau, \s$ and we choose
 $\tau=t$ and $\s = y$ for convenience. 
Since $\CT\gg L$, we can assume that the surface is translationally invariant along $\tau$ direction and therefore $X^{M}(\s)$ depends on $\s$ only. 
The ``Lagrangian" in \eqref{S_def} reads: 
\be
\label{L}
\CL (r',z';r)= \sqrt{A(r)\[1+ \(B(r)\)^{-1}(r')^2\] + C(r)(z')^2 +2 \vanom D(r) z' r'}\, ,
\ee
where $A(r), B(r), C(r)$, to linear order in $\vanom$,
are given by:
\begin{eqnarray}
\label{ABCD}
A(r) &=& 
\cosh^2\eta \, g_{xx}^2(r)\, \[ f(r)-\tanh^2\eta - 2 \vanom\tanh\eta\, Q(r)\] 
+\CO(\vanom^2)
\, , 
\no \\
B(r)&=& 
f(r) g_{xx}^2(r) +\CO(\vanom^2)\, , 
\qquad
C(r)= \cosh^2\eta\, g^2_{xx}(r) f(r)+\CO (\vanom^2)\, ,
\no \\
D(r) &=& -\cosh\eta\, \gxx Q(r)+\CO(\vanom)
\, . 
\end{eqnarray}
Here and hereafter, we use prime to denote the derivative with respect to $\s$ (or equivalently to $y$). 
As $\CL(r',z'; r)$ is independent of $z$, the momentum $\pi_{z}$ conjugated to $z'$ is a constant of the motion:
\bes
\label{COM}
\be
\label{piz}
\pi_{z} \equiv \frac{\pd \CL}{\pd z'} = \frac{C(r)z' + D(r)r'}{\CL}=\text{const}\, . 
\ee 
Moreover, 
as $\CL(r',z'; r)$ has no explicit dependence on $\s$, 
the corresponding Hamiltonian $H$:
\be
\label{H}
H \equiv \CL - r' \frac{\pd \CL}{\pd r'} - z' \frac{\pd \CL}{\pd z'}
= \frac{A(r)}{\CL}
= \text{const}
\, ,
\ee
\ees
is also a constant of motion. 
From \eqref{COM} and boundary conditions,
\be
\label{bc}
z\left(\frac{L}{2}\right)=z\left(-\frac{L}{2}\right)=0\, , 
\qquad
r\left(\frac{L}{2}\right) = r\left(-\frac{L}{2 }\right) = \Lambda \to \infty \, ,
\ee
$r(\s)$ and $z(\s)$ can be determined for each given set of integration constants.
Here $\Lambda$ is a cut-off along holographic direction. 
We will show in Sec.~\ref{sec:z1} that $\pi_{z}=0$ for the given setup. 
Therefore the solution only depends on the integration constant $H$.  In fact, for a given $H$, one could determine $L$ by solving \eqref{COM} and \eqref{bc}. Consequently, one might consider $L(H)$ as a function of $H$. Generically, $L(H)$ will reach a maximum at some $H$, say $H_{m}$ (see for example Ref.~\cite{Liu:2006nn}). 
Following Refs.~\cite{Liu:2006nn}, 
we will interpret:
\be
l_{s} \equiv L(H_{m})\, , 
\ee
as the screening length of the quarkonium potential. 
The physical picture of such definition of the screening length is clear: for $L>l_{s}$, 
\eqref{COM} has no solutions.
Therefore there is no $L$-dependent potential between the quark and antiquark 
and attractive force between a quark and anti-quark pair separated by $L>l_{s}$ will be screened. 
 
\section{The screening length $l_{s}$ in the anomalous flow}
\label{sec:ls}

In this section, 
we will determine the anomalous contribution at the linear order in $\vanom$. 
In particular,
we will expand $L(H;\eta)$ and $l_{s}(\eta)$ as:
\begin{eqnarray}
\label{L_expand}
&~&L(H;\eta) = L_{0}(H;\eta)+\vanom L_{1}(H;\eta)+\CO(\vanom^2)\, ,\notag\\
&~&l_{s}(\eta;\vanom)=l_{0}(\eta)+ \vanom l_{1}(\eta)+\CO(\vanom^2)\, ,  
\end{eqnarray}
and obtain expressions for $L_{1}(H;\eta)$ and $l_{1}(\eta)$.
The zeroth order results (i.e. results in the absence of anomalous effects) such as string profile
 $r_{0}(\s), z_{0}(\s)$ and 
$l_{0}(\eta)$
 are extensively discussed in literature~\cite{Liu:2006nn,Caceres:2006ta}. 
Using the results of Ref.~\cite{Liu:2006nn},
we have (see also Sec.~\ref{sec:z1}):
\be
\label{z0}
z_{0}(\s)=\text{const}\, . 
\ee
Eq.~\eqref{z0} implies that $z'\sim \vanom$ and
hence \eqref{H} becomes
\be
H = \frac{A(r)}{ \sqrt{A(r)\[1+ (r')^2/B(r)\]}} + \CO(\vanom^2)\, . 
\ee
This leads to the expression:
\be
\label{r_s}
\frac{d r}{d\s} = - \frac{1}{H}\sqrt{B(r)\[ A(r) - H^2\]}\, . 
\ee
We note that with given $H$,
 R.H.S of \eqref{r_s} will vanish at $r=r_{c}$ with $r_{c}$ satisfying:
\be
\label{rc_def}
A(r_{c}) = H^2\, . 
\ee
Due to $r'(\s)=0$ at this point, $r_{c}$ would be the minimum $r$ the world sheet would reach. 
In other words, the world sheet stretches from $r=\Lambda\to \infty$ down to $r=r_{c}$. 
Since $r(\s)$ is an even function of $\s$, 
using boundary condition \eqref{bc},
we have:
\be
\label{L_fun}
L(H;\eta) =2H \int^{\Lambda}_{r_{c}} dr \frac{1}
{
\sqrt{B(r)\[A(r)-H^2\]}
}  
\, . 
\ee

It is clear from the above derivation that zeroth order results $L_{0}(H)$ can be recovered by the replacement $A(r)\to A_{0}(r),~r_{c}\to r^{0}_{c}$ in \eqref{L_fun}:
\be
\label{L_fun1}
L_{0}(H;\eta) =2H \int^{\Lambda}_{r^{0}_{c}} dr \frac{1}
{
\sqrt{B(r)\[A_{0}(r)-H^2\]}
}  
\, , 
\ee
where $r^{0}_{c}$ is fixed by: 
\be
A_{0}(r^{0}_{c}) = H^2\, . 
\ee
Here we define $A_{0}$ and $A_{1}$ by expanding $A$ in powers of $\vanom$: $A(r)=A_{0} (r) + \vanom A_{1}(r)$ and one finds using \eqref{ABCD} that 
\begin{eqnarray}
\label{eq:A_expand}
A_{0}(r) =\cosh^2\eta\, g^2_{xx}(r)\[f(r) -\tanh^2\eta\]\, ,  
\qquad
A_{1}(r) =-\sinh(2\eta) g^{2}_{xx}(r)Q(r)\, . 
\end{eqnarray}
As the anomalous contribution to $B$ is already at $\CO (\vanom^2)$ , 
we will not distinguish $B$ from its zeroth order part $B_{0}$
(similar for $C$ and $D$ used in Sec.~\ref{sec:z1} ).

At the linear order in $\vanom$, one can determine $L_{1}(H;\eta)$ defined in \eqref{L_expand} by expanding \eqref{L_fun} in powers of $\vanom$.
As a result (see Appendix.~\ref{sec:derivation} for details),
we have:
\begin{eqnarray}
\label{LvsH}
 L_{1}(H;\eta)
= -\frac{2 H  A_{1}(\rca)}{V_{A}(\rca)}\< \vAfun;H\>\, , 
\end{eqnarray} 
where we introduce
\begin{equation}
\label{vA_fun}
\vAfun \equiv \frac{\vanom Q(r)}{1-f(r)}\, .
\end{equation}
We will discuss the physical interpretation of $\vAfun$ shortly.
In \eqref{LvsH},
we have defined the ``average" over holographic coordinate for any function of $F(r)$:
%
%
\begin{eqnarray}
\label{average}
\<F(r);H\>
&\equiv&
\int^{\infty}_{\rca}dr\, 
\Bigg\{\[\frac{\(1-f(r)\)g^2_{xx}(r)}{\(1-f(\rca)\)g^2_{xx}(\rca)}F(r)
-\frac{A'_{0}(r)r}{A'_{0}(\rca)\rca}F(\rca)\]\times \frac{1}{ 2\[B(r)\]^{1/2}\[A_{0}(r)-A_{0}(\rca)\]^{3/2}}
\no \\
&+&
\frac{F(\rca)}{A'_{0}(\rca)\rca}\Big[\frac{2 B(r)
-B'(r)r}{2\[B(r)\]^{3/2}\[A_{0}(r)-A_{0}(\rca)\]^{1/2}}\Big]
\Bigg\}\, ,
\end{eqnarray}
to make our notations compact. 
Obviously, due to \eqref{vA_fun}, $L_{1}(H;\eta)$ determined from \eqref{LvsH} is independent of $\vanom$.

With \eqref{LvsH}, we now ready to determine $l_{s}(\eta)$. 
Let $H^{0}_{m}$ and $H_{m}$ be the $H$ corresponding to the maximum of $L_{0}(H;\eta)$ and $L(H;\eta)$ respectively, i.e. $l_{0}(\eta)=L_{0}(H^{0}_{m};\eta),~l_{s}(\eta)=L(H_{m};\eta) $. 
Note that $L_{0}(H_{m};\eta)=L_{0}(H^{0}_{m};\eta) +\CO(\vanom^2)$ due to $\pd_{H}L_{0}(H^0_{m};\eta)=0$. 
Hence we have $L(H_{m};\eta)= L_{0}(H^0_{m};\eta) +\vanom L_{1}(H_{m};\eta)=L_{0}(H^0_{m};\eta) +\vanom L_{1}(H^{0}_{m};\eta)+O(v_A^2)$ where we noticed that the difference between $H^{0}_{m}$ and $H_{m}$ is of the order $\vanom$. 
We therefore have 
\begin{equation}
\label{ls_relation}
l_{s}(\eta)= l_{0}(\eta)+\vanom L_{1}(H^{0}_{m}(\eta);\eta)\, , 
\end{equation}

Now one can compute $\etaA$,
the ``anomalous flow" felt by the charmonium moving in a chiral plasma using \eqref{eta_anom}.
From \eqref{ls_relation},
we easaily find:
\begin{eqnarray}
\label{va_shift}
\etaA
&=&  
\frac{\<\vAfun;H^{m}_{0}\>}{\<1 ;H^{m}_{0}\>}\, . 
\end{eqnarray} 
Eq.~\eqref{va_shift} is one of the main results of this paper. 
We note from \eqref{average} and \eqref{ABCD} that $A_{0}, B$ is invariant under $\eta \to -\eta$, therefore $\eta_{\anom}$ is an even function of $\eta$. 

To analyze \eqref{va_shift},
we first discuss the physical meaning of $\vAfun$ defined in \eqref{vA_fun}.
We note that the chiral anomaly will induce a non-zero $g_{tz}(r)=\gxx Q(r)$ in the metric \eqref{metric2}. 
The presence of this $g_{tz}(r)$ in the metric is equivalent to a boost of the non-anomalous metric with $r$-dependent velocity $\vAfun$ (keeping terms only up to the linear order in $\vanom$):
\begin{eqnarray}
\label{va_boost}
dt= dt' - V_A(r) dz'\, , 
\qquad
dz=- V_A(r) dt' + dz'
\, .
\end{eqnarray}
Therefore the bulk velocity $\vAfun$ describes an ``anomalous wind" flowing in the bulk. The ``anomalous flow" $\etaA$ probed by a charmonium is given by an average of $\vAfun$ over the holographic coordinate $r$ with an appropriate weight as given by \eqref{average}. 
It is clear that the resulting $\etaA$ will depend on the details of the bulk profiles thus will be {\it non-universal}. 

\begin{figure}
\centering
\includegraphics[width=0.45\textwidth]
{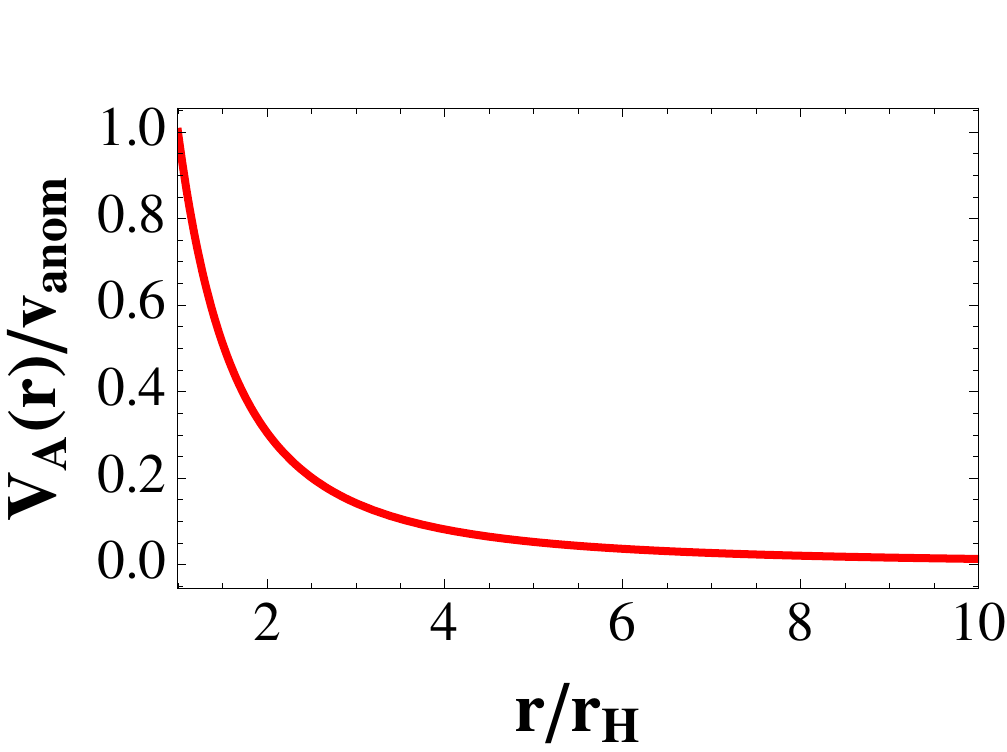}
\caption
{
\label{fig:VA}
$V_A(r)/\vanom$,
the bulk ``anomalous velocity",  introduced in \eqref{vA_fun} as a function of $r/r_{H}$ for ${\cal N} =4$ SYM chiral plasma at finite chemical potential 
(c.f.~Sec.~\ref{sec:results}).
In numerical computations, 
we consider right-handed fermions with $\mu_R/T=0.1$ (to the leading order in $\mu_R/T$).
}
\end{figure}
In Fig.~\ref{fig:VA},
we plot a representative $\vAfun/\vanom$ vs $r$ for ${\cal N} =4$ SYM chiral plasma at finite chemical potential. 
Note that the asymptotics of $\vAfun$ is model-independent. 
As one can check, 
the energy flow of the chiral fluid dual to \eqref{metric2} is proportional to $V_{A}(r\to\infty)$. 
As we are working in the Landau frame, the energy flow  $V_{A}(r\to\infty)$ must vanish. 
On the other hand,
at the horizon $r\to r_{H}$ one finds $V_{A}(r\to r_{H})\to \vanom$. 
This is because $V_{A}(r\to r_{H})$ is related to the entropy flow \eqref{va_def} induced by the anomaly which is model-independent. 
Indeed, 
As Fig.~\ref{fig:VA} indicates,
$\vAfun$ decreases from $\vanom$ to $0$ from boundary towards the horizon.  

\subsection{String stretched by ``anomalous flow"}
\label{sec:z1}
As another manifestation of the ``anomalous flow" experienced by a moving charmonium,
we now consider string profile along $z$-direction $z(\s)$.
As mentioned before, we expand string profile $z(\s)$ as $z(\s)=z_{0}(\s)+\vanom z_{1}(\s)$. 

First, 
let us briefly recall the derivation of \eqref{z0} that $z_{0}(\s)=0$ following the argument presented in Ref.~\cite{Liu:2006nn}. 
In the absence of the anomaly, 
\eqref{piz} becomes
\begin{equation}
\label{pi0}
\pi^{0}_{z}
= \frac{C(r)z'}{\CL}\, . 
\end{equation}
If $\pi^0_z$ is non-zero then $z'_0\neq0$. The boundary condition \eqref{bc} implies that there is at least a point at which $z'_0=0$. 
 On the other hand, the Lagrangian $\CL$ has be to positive definite. 
 Therefore R.H.S of \eqref{z0} vanishes at this point where $z'=0$. That is in contradiction with the assumption that $\pi^{0}_{z}\neq 0$ and thus we conclude that $\pi_z^0=0$ and $z_{0}(\s)={\textrm const}$.
 
\begin{figure}
\centering
\includegraphics[width=0.5\textwidth]
{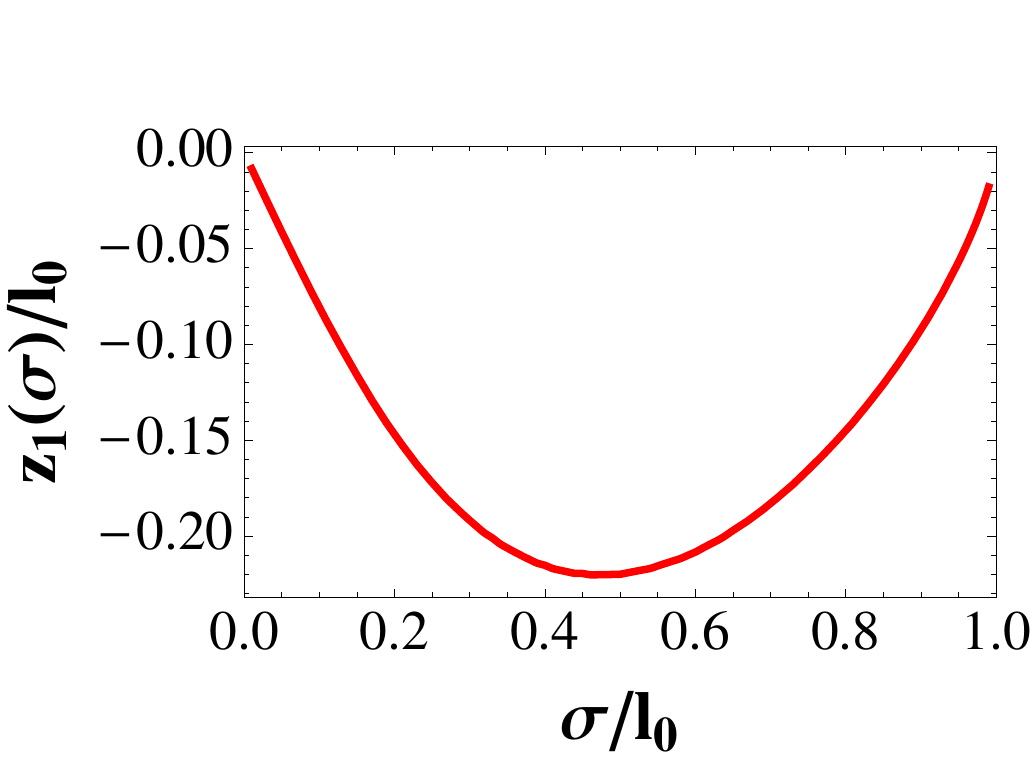}
\caption
{
\label{fig:string}
(Color Online) 
The anomalous contribution to the string profile $z_1(\sigma)$ as given by \eqref{z1_sol} for ${\cal N} =4$ SYM chiral plasma at finite chemical potential 
(c.f.~Sec.~\ref{sec:results}).
The string is ``dragged" by the anomalous flow. 
Both $z_{1}(\s)$ and $\s$ are rescaled by the corresponding zeroth order screening length $l_{0}$. 
In numerical computations, 
we consider right-handed fermions only with $\mu_R/T=0.1$ and $v=0.5$ (to the leading order in $\mu_R/T$).
}
\end{figure}

The situation is different due to anomaly. The condition \eqref{piz} at the leading order in $\vanom$ can be expressed in terms of the zeroth-order profile:
 \bes
 \be
 \label{pi1}
 \pi^{1}_{z} = \frac{C\[r_{0}(\s)\] z'_{1}(\s)+ D\[r_{0}(\s)\]r'_{0}(\s)}{\sqrt{A_{0}\[r_{0}(\s)\]\{1 + B_{0}\[r_{0}(\s)\]\}\[r_{0}'(\s)\]^2}}= 
 \frac{H}{A_{0}\[r_{0}(\s)\]} \{ C\[r_{0}(\s)\] z'_{1}(\s)+  D\[r_{0}(\s)\]r'_{0}(\s) \}\, ,
 \ee
 where we have used the zeroth order relation for \eqref{H}:
 \be
 \label{H1}
H = \frac{A_{0}(r)}{ \sqrt{A_{0}(r)\[1+ B(r)(r')^2\]}}\, . 
\ee
\ees
Solving \eqref{pi1} and the using boundary condition $z_{1}(-L/2)=0$, we then have:
\begin{equation}
\label{z1_sol0}
z_{1}(\s)
=
\(\frac{\pi^{1}_{z}}{H}\)\int^{\s}_{-L/2}d\tilde\s\, 
\frac{A_{0}\[r_{0}(\tilde\s)\]}{C\[r_{0}(\tilde\s)\]}
- \int^{\s}_{-L/2}d\tilde\s\frac{D\[r_{0}(\tilde\s)\]}{C\[r_{0}(\tilde\s)\]}
r'_{0}(\tilde\s) \, . 
\end{equation}
The integration constant $\pi^{1}_{z}$ can be fixed by taking $\s\to L/2$ in \eqref{z1_sol0} and imposing the boundary condition $z_{1}(L/2)=0$. 
Noting $r_{0}(\s)$ is an even function of $\s$ and $r'_{0}(\s)$ is an odd function of $\s$, we find:
\begin{equation}
\label{zL}
z_{1}(L/2)
=
\(\frac{\pi^{1}_{z}}{H}\)\int^{L/2}_{-L/2}d\tilde\s\, 
\frac{A_{0}\[r_{0}(\tilde\s)\]}{C\[r_{0}(\tilde\s)\]}\, . 
\end{equation}
The integration in R.H.S of \eqref{zL} is non-zero since $A_0\(r_0(\sigma)\)/C\(r_0(\sigma)\)$ is even function of $\sigma$. Thus to satisfy boundary conditions,
we must have also $\pi^{1}_{z}=0$ and consequently \eqref{z1_sol0} becomes:
\be
\label{z1_sol}
z_{1}(\s)
=-  \int^{\s}_{-L/2} d\tilde{\s} \frac{D\(r_0\(\tilde{\s}\)\)}{C\(r_0(\tilde{\s})\)} r'_0(\tilde{\s})\, . 
\ee
A representative profile of $z_{1}(\s)$ is plotted in Fig.~\ref{fig:string} . 
As one can check from \eqref{z1_sol}, 
$z_{1}(\s)$ is negative. 
This implies that the string profile along $z$-direction is dragged behind the ``anomalous wind" $v_{\anom}$. 
From \eqref{ABCD},
we note $C$ and $D$ are invariant under $\eta \to -\eta$, therefore $z_{1}$ is an even function of $\eta$. 


Let us contrast the anomalous contribution \eqref{z1_sol} with zero order results \eqref{z0}.
$z_{0}(\s)=0$ suggests that even though there is a wind blowing in the $z$-direction, 
 the string world sheet is not dragged at all by this wind. 
However, 
non-trivial $z_{1}(\s)$ given by \eqref{z1_sol} indicates that string will be stretched by the anomalous flow. 
It would be interesting to gain further physical insights in this aspect. 

\section{Representative results for ${\cal N}=4$ SYM chiral plasma}
\label{sec:results}
Up to this point, our discussion is general and applicable to any metric of the form \eqref{metric2}. 
We now take ${\cal N}=4$ SYM theory as an example and report  the anomalous contribution to color screening length as defined in \eqref{L_expand} .

For ${\cal N}=4$ SYM theory with a nonzero chemical potential and with a weak homogeneous external magnetic field $\vB$, we have (see e.g. \cite{Megias:2013joa}):
\begin{eqnarray}
\label{fandQ}
\gxx &=&r^2\, ,
\qquad
f(r)=1- \frac{M}{r^4}+ \frac{q^2}{r^6}\, ,
\no \\ 
Q(r)&=&\frac{2\pi^2T^2}{r^2}-\frac{\pi^4T^4}{r^4}+2\left(1-\frac{\pi^4 T^4}{r^4}\right)\log\left(\frac{r^2}{r^2+\pi^2T^2}\right)+O(\mu^2/T^2)\, ,
\end{eqnarray}
where for brevity we kept only the leading term in $Q(r)$ in powers of $\mu/T$.
In this section, 
we will consider one chirality, say right handed fermions, only and therefore $\mu$ should be interpreted as a chiral chemical potential $\mu_{R}$. 
This is sufficient for our illustrative purpose. 

Black hole parameters $M,~q$ could be easily related to physical quantities $\mu$ and $T$ (see e.g. \cite{Erdmenger:2008rm, Rajagopal:2015roa}):
\begin{equation}
M =\frac{\pi^4T^4}{16}\[\sqrt{1+\frac{8\mu^2}{3\pi^2T^2}}+1\]^{3}
\[3\sqrt{\frac{8\mu^2}{3\pi^2T^2}+1}-1\]\, , 
\qquad
q=\frac{\mu}{\sqrt{3}}\frac{\pi^2T^2}{2}\[\sqrt{1+\frac{8\mu^2}{3\pi^2T^2}}+1\]^{2}\, .
\end{equation}
As we will measure any dimensionful quantities in units of $\pi T$,
 the results then will depend on  the ratio $\mu/T$.
In numerics,
we choose a specific value $\mu/T=0.1$ and use $Q(r)$ and $f(r)$ given by \eqref{fandQ}. 
\begin{figure}
\centering
\subfigure[]
{
\label{fig:eta_anom}
\includegraphics[width=0.45\textwidth]{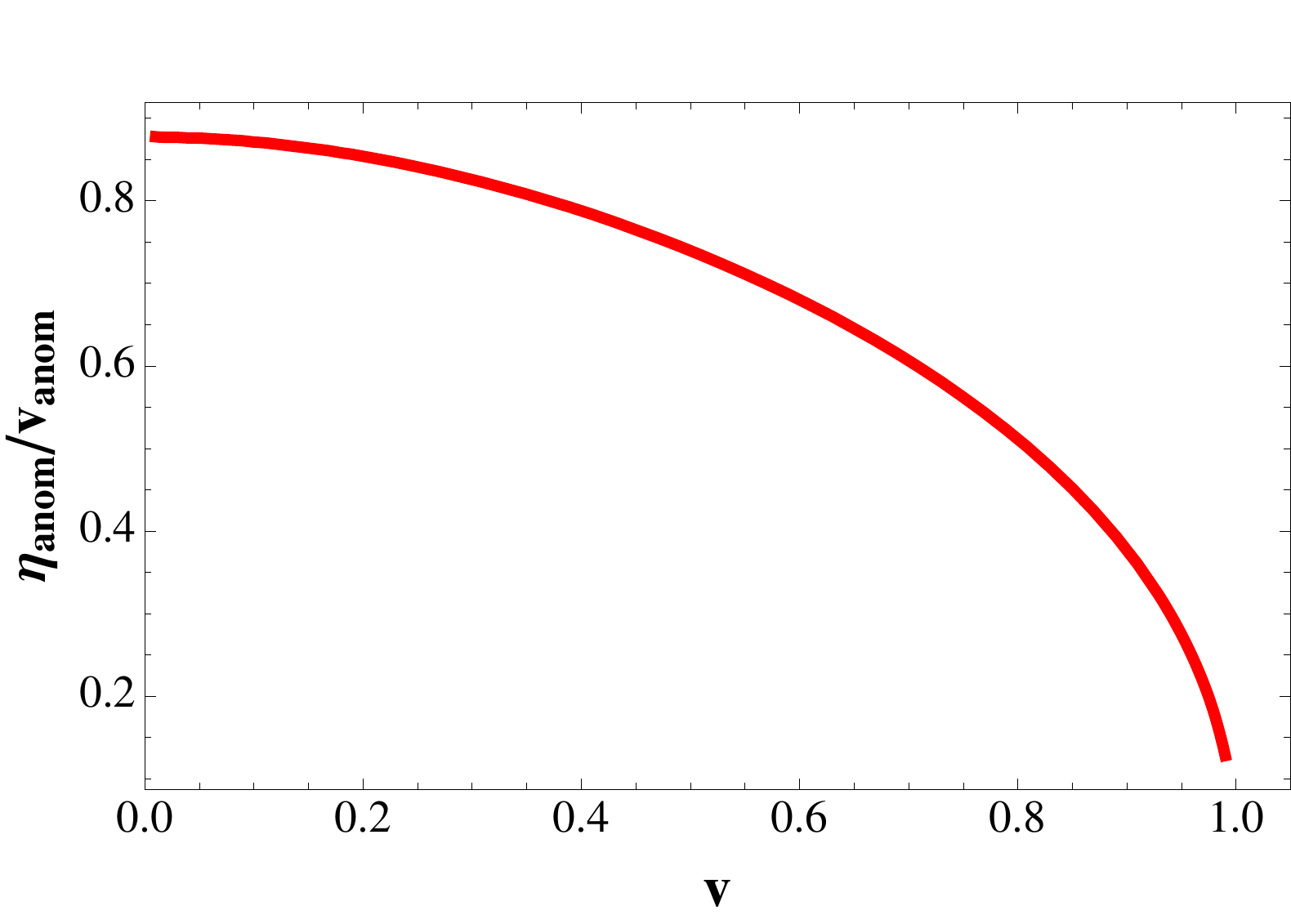}
}
\subfigure[]
{
\label{fig:dls}
\includegraphics[width=0.45\textwidth]{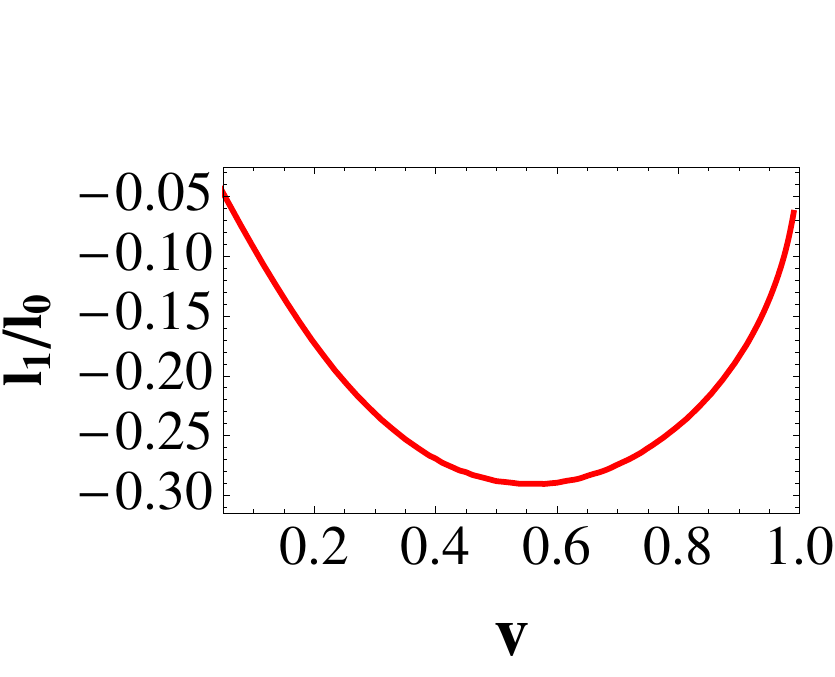}
} 
\caption
{
Left: 
anomalous velocity $\eta_{\anom}$ ``felt" by a charmonium moving at velocity $v$ in a chiral plasma with a anomalous entropy current $s\, \vanom$. 
Here, $\eta_{\anom}$ is determined from \eqref{va_shift} and is along the direction of anomalous entropy current in Landau frame.
Right: anomalous contribution to the color screening length $l_s$. 
As in \eqref{L_expand}, 
we expand the color screening length as $l_{s}=l_{0}+\vanom l_{1}+\CO (\vanom^2)$.
In numerical computations, 
we consider ${\cal N} =4$ SYM chiral plasma at finite chemical potential with right-handed fermions only with $\mu_R/T=0.1$ (to the leading order in $\mu_R/T$).
We have also translated the dependence on rapidity $\eta$ into that of velocity $v$. 
}
\end{figure}

In Fig.~\ref{fig:eta_anom},
we plot $\eta_{\anom}$ vs $v$ as determined from \eqref{va_shift}. 
Observe as expected that $\eta_{\anom}$ is along the direction of the entropy flow $\vanom$. 
With increasing $v$, 
$\eta_{\anom}$ will decreases towards zero. 
This is of course not surprising	 as a ultra-relativistically moving charmonium should be insensitive to a small flow $\vanom$. 
Furthermore, 
In Fig.~\ref{fig:dls}, 
we plot the (relative) anomalous contribution to the screening length, $l_{1}(v)/l_{0}(v)$ as a function of velocity $v$.
Here, $l_{0}$ and $l_{1}$ are defined in \eqref{L_expand} by expanding the color screening length in the presence of the anomaly to the leading order in $\vanom$. 
We observe that $l_{1}(v)/l_{0}(v)$ reaches its maximum at some intermediate velocity $v$. 
For small $v$, $l_{s}(v)/l_{0}(v)$ grows linearly in $v$. 
At high velocity $v$, 
$ l_{1}(v)/l_{0}(v)$ is suppressed due to the vanishing of $\eta_{\rm anom}$ as shown in Fig.~\ref{fig:eta_anom}. 
It is worthy noting that strictly speaking, 
in our small $\vanom$ expansion, 
we have in-explicitly assumed that $\vanom\ll v$.
Therefore small $v$ asymptotic should be understood as taking $v\to 0$ limit while keeping $\vanom \ll v$. 

\section{Discussion and Phenomenological Consequences}
\label{sec:summary}

In this work, 
we consider the charmonium color screening length $l_{s}$ in a strongly coupled chiral plasma in the presence of flow induced by chiral anomaly. 
We use holographic correspondence to show $l_{s}$ is influenced by the anomalous flow and establish an analytical formula \eqref{va_shift}
to quantify such influence. 
From the gravity side of the correspondence,
such contributions stems from the modification of the bulk metric due to the chiral anomaly (c.f.~\eqref{metric2}).
Consequently, it contributes to Nambu-Goto action \eqref{S_def} which determines the color screening length.  
While the anomalous effects on light quarks which are nearly chiral have been studied extensively, 
exploring the anomalous plasma using heavy probes is initiated very recently. 
Our results
as a proof of principle,
 imply that those anomalous effects can be probed by charmonium. 
We will further discuss the physical interpretation and possible phenomenological implication below. 

\subsection{Comparison with anomalous contribution to heavy quark drag force}
It is instructively to compare our results with the recent study of Ref.~\cite{Rajagopal:2015roa} in which drag force $F(\eta;\vanom)$ of a heavy quark moving at rapidity $\eta$ in the presence of anomalous velocity $\vanom$ (c.f.~\eqref{va_def}) is studied. 
Let us rephrase the results in Ref.~\cite{Rajagopal:2015roa} by introducing $\eta^{\drag}_{\anom}$ which describes the ``anomalous flow" experienced by a heavy quark.
Similar to \eqref{eta_anom} where the ``anomalous flow" experienced by a charmonium $\eta_{\anom}$ is introduced,
we define $\eta^{\drag}_{\anom}$ by the condition:
\begin{equation}
\label{F_def}
F(\eta;\vanom)=F(\eta+\eta^{\text{drag}}_{\anom};\vanom=0)\, . 
\end{equation}
Ref.~\cite{Rajagopal:2015roa} indicates that $\eta_{\anom}$ is non-zero and is proportional to $\vanom$ as well. 

However, 
one should not overlook the {\it qualitative} difference between the case of the color screening length presented in here and that of the heavy quark drag coefficient.
In fact, 
$\eta^{\drag}_{\rm anom}$ is independent of the microscopic details of the system
while $\eta_{\anom}$, 
as shown in this paper, 
is model-dependent. 
For example, 
in Landau frame and in $\eta\to 0$ limit (i.e the heavy probe is at rest), 
one can see from Ref.~\cite{Rajagopal:2015roa} that $\eta^{\drag}_{\rm anom}=V_{A}(r_{H})$.
Therefore $\eta^{\drag}_{\anom}$ only depends on the value of $\vAfun$ at horizon and is uniquely fixed by $\vanom$.
In contrast, 
$\etaA$ in \eqref{va_shift} is given by the bulk average over $\vAfun$ and therefore depends on the bulk details of the holographic model. 
This remarkable difference is directly connected to the dissipation-less nature of the anomalous transport. 
 As the presence of the drag force introduces dissipation,
 there should be a frame for a chiral plasma with anomalous flow that drag force vanishes. 
In fact $-(\eta+\eta^{\drag}_{\anom})$ defines that frame as $F(0;\vanom=0)=0$. 
Therefore $\eta_{\anom}$ is constrained by the dissipation-less nature of the anomalous transport. 
 Indeed, 
 $\eta_{\anom}$ can be universally fixed by using anomalous hydrodynamics in Ref.~\cite{Stephanov:2015roa} and by the generalization of Landau's criterion for superfluidity in Ref.~\cite{Sadofyev:2015tmb}. 
On the other hand,
 the color screening length is a static quantity and is not related to dissipative processes. 
This in turn implies that one can not constrain anomalous contributions to $l_{s}$ directly from argument based on non-dissipationless nature of chiral effects.
Our holographic study confirmed this difference.  

\subsection{Phenomenological implications}
We now turn to phenomenological implications. 
As the anomalous contributions is controlled by the magnitude of the ``anomalous flow" $\vanom$,
let us begin our discussion by estimating the magnitude of $\vanom$ in heavy-ion collisions. 
Recovering dependence on charge $e$, $C_{A}$ in \eqref{va_def} we get:
\be
\label{CA_QCD}
C_{A}= \frac{C_{\rm EM}e^2}{2\pi^2}\, . 
\ee
We consider the case that only $u,d$ will contribute to the CME current
therefore $N_{f}=2, C_{\rm EM}=5/9$. 
To estimate $\e$ and $p$,
we use the equation of state of a massless ideal quark-gluon gas
$
\varepsilon = 3p
$, 
with 
\begin{equation}
p\left(
T
\right) = 
\frac{g_{\rm QGP} \pi^2}{90 }T^4 \, , 
\end{equation}
where $g_{\rm QGP} = g_{g} + 7g_{q}/8$ is the number of
degrees of freedom with $g_{g} = (N_c^2 - 1) N_s$ and $g_{q} = 2 N_c N_s
N_f$; $N_c = 3$ and $N_f = 2$ are the numbers of colors and flavors
and $N_s =2$ is the number of spin states for quarks and (transverse)
gluons. 

As a results, we have:
\begin{equation}
  \label{eq:1}
 \vanom
\approx 0.003 \(\frac{eB}{T^2}\)\(\frac{\mu_{V}\mu_{A}}{T^2}\)\, . 
\end{equation}
Keeping  in mind the characteristic $\sqrt{eB}$ and $\mu $ (is of order $T$) one can see that $\vanom$ is numerically small.
This implies that our results based on small $\vanom$ expansion is applicable to many situations in heavy-ion collisions. However anomalous effects on charmonium dissociation are numerically small. 

Moreover, 
we note that at early time of heavy-ion collisions, 
$eB$ can be of the order $10~m^2_{\pi}$ at RHIC and of the order $100~m^2_{\pi}$ at LHC. 
Those numbers can be even larger from event by event fluctuations~\cite{Bzdak:2011yy,Deng:2012pc}. 
This would lead to a significant $\vanom$ .
Therefore it would be interesting to consider the situation that $B$ is large and 
explore if anomaly-induced $\eta_{\anom}$ would lead to a new mechanism for charmonium suppression at early time of heavy-ion collisions.  
We hope the results presented in this paper would encourage further study along this direction. 

The current work can be extended in number of ways.
In this exploratory study,
we consider charmonium moving along the direction of anomalous flow and take the configuration of the heavy quark ``dipole" to be perpendicular to the anomalous flow. 
The study for general angle between the anomalous flow and the ``dipole" velocity would bring further details on anomalous contributions to the screening length. 
We have restricted ourselves to the anomalous flow related to the chiral magnetic effects.
One could extend it by incorporating chiral vortical effects ~\cite{Banerjee:2008th,Erdmenger:2008rm,Kirilin:2012mw} and possible contribution from gravitational anomaly~\cite{Landsteiner:2011cp}.
We also assume both the magnetic field and axial charge to be static and homogeneous. 
In realistic situations such as the early times of heavy-ion collisions, 
both magnetic field and axial charge imbalance are dynamical. 
Those dynamical magnetic field and axial charge imbalance are shown to induce novel phenomena related to chiral anomaly (see for example Ref.~\cite{Khaidukov:2013sja,Buividovich:2013hza,Iatrakis:2014dka,Iatrakis:2015fma,Hirono:2015rla}).
It would be interesting to understand their influences on the color screening. In the realistic situation the plasma is also anisotropic. 
Both the color screening length and the anomalous transport (see e.g. \cite{Gahramanov:2012wz}) depend on the anisotropy of the medium (see e.g. \cite{Chernicoff:2012bu}). 
Thus one would expect some interplay between them.
However these directions are beyond the scope of this work and we leave them for future study. 


\acknowledgments
We would like to thank Krishna Rajagopal and Ho-Ung Yee for helpful discussions.
This work was supported in part by DOE Contract No.~DE-SC0011090 (AS) and in part by Contract No. DE-SC0012704 (YY). AS is grateful for travel support from RFBR grant 14-02-01185A.

\begin{appendix}

\section{Derivation of \eqref{LvsH}}
\label{sec:derivation}
We now present the derivation of \eqref{LvsH}.
From \eqref{L_fun} and \eqref{L_fun1}, 
we have:
\begin{eqnarray}
\label{DL}
\vanom L_{1}(\eta)
&=& L(H;\eta) - L_{0}(H;\eta) 
\no \\
&=& 2 H \{  
\int^{r}_{r_{c}} dr \frac{1}{\sqrt{B(r)\[ A(r)- H^2\]}}
- 
\int^{r}_{r^{0}_{c}} dr \frac{1}{\sqrt{B(r)\[ A_{0}(r)- H^2\]}}
\}  
\no \\
&=&
2 H \{ 
\int^{\infty}_{1} d\tr \frac{r_{c}}{\sqrt{B(r_{c}\tr)\[ A(r_{c}\tr)- H^2\]}} -
\int^{\infty}_{1} d\tr \frac{r^{0}_{c}}{\sqrt{B(r^{0}_{c}\tr)\[ A_{0}(r^{0}_{c}\tr)- H^2\]}}
\}
\no \\
&=&
2 H 
\int^{\infty}_{1} d\tr 
\{ \frac{r_{c}}{\sqrt{B(r_{c}\tr)\[ A(r_{c}\tr)- H^2\]}} -
\frac{r^{0}_{c}}{\sqrt{B(r^{0}_{c}\tr)\[ A_{0}(r^{0}_{c}\tr)- H^2\]}}
\}\, . 
\end{eqnarray}
In \eqref{DL},
we have introduced a rescaled variable $\tr$ such that the integration is from $1$ to $\infty$. 
This would bring the convenience to expand \eqref{DL} in power of $\vanom$,

Let us now expand: 
\be
 r_{c} = r^{0}_{c} + \vanom r^{1}_{c}+
+ \CO (\vanom^2)\, ,
\ee
to obtain:
\begin{eqnarray}
\label{AE_expand}
A(r_{c}\tr)
&=&  A_{0}(r^{0}\tr) + \vanom 
\Big[A_{1}(r^{0}_{c}\tr)+A'_{0}(r^{0}_{c} \tr) \(r^{1}_{c} \tr \)\Big ]
+\CO (\vanom^2) \, , 
\no \\
B(r_{c}\tr)
&=& B(r^{0}_{c}\tr)  + \vanom B'(r^{0}_{c} \tr) \(r^{1}_{c} \tr \) +\CO (\vanom^2)
\, . 
\end{eqnarray}
In the above expansion \eqref{AE_expand},
we have assumed that $\(r A'_{0}(r)\)/A_{0}(r)$ and $\(r B'(r)\)/B(r)$ are finite for arbitrary $r$.
As one can check, the is indeed the case for asymptotic AdS metric. 

Substituting \eqref{AE_expand} into \eqref{DL} and comparing terms proportional to $\vanom$ in \eqref{DL},
we have:
\be
\label{DL1}
 L_{1}(\eta)
= 2H
\int ^{\infty}_{r_c^0}d r 
\{
- \frac{\[ A_{1}(r) + A'_{0}(r)\( r^{1}_{c}/r^{0}_{c}\)r \]}{2 \[B(r)\]^{1/2}\[ A_{0}(r) - A_{0}(r^0_{c})\]^{3/2}}
+\(\frac{r^{1}_{c}}{r^{0}_{c}}\)
\[\frac{2 B(r)- B'(r)r}{2  \[B(r)\]^{3/2}\[ A_{0}(r) - A_{0}(r^0_{c})\]^{1/2}} \]
\}\, ,
\ee
where we have replaced $\tr$ with $r/r^{0}_{c}$ and $H^2$ with $A_{0}(r^0_{c})$.
Finally,
from the definition of $r_{c}$ ($r^{0}_{c}$) \eqref{rc_def}, 
i.e., $A(r_{c})=A_{0}(r^{0}_{c})=H^2$, 
we obtain:
\be
\label{eq:Dr0}
 r^{1}_{c} = - \frac{A_{1}(r^{0}_{c})}{A'_{0}(r^{0}_{c})}\, . 
\ee
Putting \eqref{DL1} and \eqref{eq:Dr0} together, we arrived at \eqref{LvsH}. 
\end{appendix}

\bibliographystyle{JHEP.bst}

\bibliography{ls_and_anomaly}

\end{document}